\documentclass[9pt]{article}

\usepackage{amsmath}
\begin{document}

\title{{\bf Symmetry Classification of quasi-linear PDE's. \\ II: an 
exceptional case\\ } }
%{\large Addendum to the paper {\it Nonlinear 
%Dynamics {\bf 51}, 309--316 (2008)}}}

\author{Giampaolo Cicogna \\ ~ \\
{\it \hskip-.5cm  Dip. di Fisica ``E.Fermi'', Universit\`a di Pisa%} \\
%{\it 
\ and INFN, Sez. di Pisa 
%  Istituto Nazionale di Fisica Nucleare, Sez. di Pisa
} \\ 
 {\it Largo B. Pontecorvo 3, Ed. B-C, I-56127, Pisa, Italy} \\
(fax: +39-050-2214887; e-mail: cicogna@df.unipi.it)}

\def \ov{\over}
\def \bar{\overline}
\def \beq{\begin{equation} }
\def \eeq{\end{equation} }
\def \lb{\label}

%%%%%%% notaz matem
\def \pd{\partial}
\def\~#1{\widetilde #1}

%%%%%%%% abbrev
\def \sy {symmetry}
\def \sys {symmetries}
\def \so {solution}
\def \eq{equation}
\def \R{{\bf R}}

%%%%%%%% greco
\def\a{\alpha}
\def\be{\beta}
\def\phi{\varphi}
\def\De{\Delta}
\def\la{\lambda}
\def\ga{\gamma}

\def\Na2{\nabla^2}

\def \qq{\qquad}
\def \q{\quad} 
\def\={\, =\, }
\def\vf{vector field }
\baselineskip .5cm
\date{}
\maketitle
\def\sk{\smallskip}

\begin{abstract}
\noindent
This short note completes the symmetry analysis of a class of quasi-linear 
partial differential equations  considered in the previous paper 
(Nonlinear Dynamics {\bf 51}, 309--316 (2008)): it deals with the
presence of an ``exceptional'' Lie point symmetry, not previously 
examined, which is admitted only if the involved parameters are fixed by   
precise relationships. The peculiarity of this symmetry is enhanced 
by the fact that it leads to a  solution relevant 
in the theory of plasma physics, and also related to the presence 
of a nontrivial example of a conditional symmetry of  weak type.
\end{abstract}


\bigskip

\centerline{{\sl Running title}: Symmetry Classification II: an 
exceptional case}

%\vfill\eject
%{\twocolumn}
\bigskip \bigskip \bigskip 

%\section{The ``exceptional'' case in the symmetry\\ classification}
%\section{The ``exceptional'' symmetry and a related special solution 
%relevant in plasma physics}

This short note is an Addendum to my previous paper  \cite{ND}, and 
completes the  ``\sy\ classification'' of a class of 
quasi-linear partial differential equations. It provides the case of an
``exceptional'' symmetry, not previously considered, and a related \so\ 
which is relevant in plasma physics.

One of the examples examined in    \cite{ND} (Sect. 5) concerned \eq s of 
the form (for the unknown function $u=u(x,y)$) 
\beq\lb{genpde} 
 u_{xx}+u_{yy}+  {a \ov x}u_x\= 
 \a(x,y) \, F_1(u) +F_2(u) \q\q (a={\rm const}\not=0)
\eeq
where $\a,F_1,F_2$ are given. Equations of this type include several interesting 
cases in mathematical physics, as reduced forms of wave \eq , heat and 
Schr\"odinger \eq , and -- more specifically, as we shall see -- an 
important \eq\ in plasma physics. 

In \cite{ND}  it was been shown that 
\eq\ (\ref{genpde}) may admit Lie point \sys\ only if  both  functions $F_1\,,F_2$ have
the form
$F_i\propto u^{c_i}$ or 
$F_i\propto \exp(c_i u)$, where $c_i$ are constants.
Even in these cases, in general, only nearly 
trivial \sys\ (scalings and translations) are admitted. 

There is actually an ``exception'', not examined in 
\cite{ND}, which occurs when
\beq\lb{exc}\alpha(x,y)\=x^r\q,\q F_1(u)\=\ga_1\, u^{c_1} \q,\q\ 
F_2(u)\=\ga_2\, u^{c_2}\eeq
where $r,c_1,c_2,\ga_1,\ga_2$ are constants, and if the two following 
special relations hold 
\beq\lb{car} c_1\=1+2(r+2)/a\q\q,\q \ c_2\=1+(4/a)\ .\eeq
In this case indeed  the new nontrivial 
\sy\ generated by the \vf
\beq\lb{exsy}X\=2xy{\pd\ov{\pd x}}+(y^2-x^2){\pd\ov{\pd 
y}}-ayu{\pd\ov{\pd u}}\eeq
is admitted, as it can be easily checked.

The presence of this \sy\ implies that if 
$u=u(x,y)$ is any \so\ of Eq. (\ref{genpde}) with the choice (\ref{exc})
and (\ref{car}), then also $\~u=\~u(x,y)$ 
defined by
\beq\lb{util} \~u(x,y)\= 
[C(x,y,\la)]^{-a/2}u\big(\~x(x,y,\la),\~y(x,y,\la)\big)\eeq
where
\[C(x,y,\la)\=1+\la^2(x^2+y^2)+2\la y\]
\[\~x\=x\,[C(x,y,\la)]^{-1}\q\q\ {\rm and}\q\q\ 
\~y\=[y+\la(x^2+y^2)]\,[C(x,y,\la)]^{-1}\eqno(5')
\]
solves the \eq\ for any $\la\in\R$.  

%\section{A ``weak'' conditional symmetry and \\ a special   \so\  
%relevant in plasma physics}
\bigskip

More specifically, let me now fix $r=2$ and then consider the \eq
\beq\lb{pde2} 
u_{xx}+u_{yy}+  {a \ov x}u_x\= 
\ga_1\, x^2 \, u^{1+(8/a)} +\ga_2\,u^{1+(4/a)} 
\eeq
according to (\ref{car}). The choice $r=2$ is relevant not only because 
the resulting \eq\ has the same form as the Grad-Schl\"uter-Shafranov 
(also called 
Bragg--Hawthorne) \eq , well known in magnetohydrodynamics and plasma 
physics (see e.g. \cite{Tok} and also \cite{FOG,CPC}, in this case one has 
$a=-1$), but also because this \eq\ admits an interesting family of \so s, 
thanks to the exceptional \sy\ given above,  and to the 
following property.

It is easy to verify that the \vf
\beq\lb{Y} Y\=y{\pd\ov{\pd x}}+x{\pd\ov{\pd y}} \eeq
generates neither a standard (exact) 
\sy\ nor a (properly defined)  conditional \sy\ (see e.g.
\cite{Bl,LW,Fu,Po}) for Eq. (\ref{pde2}). Indeed, introducing as
usual the additional 
$Y$-invariance condition expressed by the first-order \eq
\beq\lb{invs}\De^{(1)}\,\equiv\,Y\,u\=0\eeq
and writing Eq. (\ref{pde2}) as $\De=0$, one has not only 
$({\tt Pr}^{(2)}Y)\De|_{\De=0}\not=0$
but also
\[({\tt Pr}^{(2)}Y)\De|_{\De=\De^{(1)}=0}\not=0\]
where as usual ${\tt Pr}^{(2)}Y$ denotes the second prolongation of  $Y$. 
However, putting $({\tt Pr}^{(2)}Y)\,\De\equiv\,\~\De$, and imposing the 
new condition $\~\De\=0$, it can be verified that
\[({\tt Pr}^{(2)}Y)\De|_{\De=\De^{(1)}=\~\De=0}=0\]
is satisfied. In this sense (see \cite{OR,CK,CL}), I say that
$Y$ is a {\it weak} conditional 
\sy\ for Eq. (\ref{pde2}). Notice that a reduction of similar type was already 
considered \cite{F1,F2},  with the name of ``anti-reduction''. 
Accordingly, a reduction procedure for Eq. (\ref{pde2}) is possible.  To this 
purpose, one can directly and more simply rewrite this \eq\  in terms 
of the $Y$-invariant variable $s=x^2-y^2$ (which solves Eq. (\ref{invs})).
Writing $v=v(s)=u(x^2-y^2)$, Eq. (\ref{pde2}) becomes
\[-4sv_{ss}+2av_s+8x^2v_{ss}\=\ga_1\,x^2\,v^{1+(8/a)} +\ga_2\,v^{1+(4/a)}\ 
.\]
The presence of the term $x^2$ in this \eq\ confirms that $Y$ is not  a 
proper conditional \sy , but the two separated \eq s 
\[-4sv_{ss}+2av_s\= \ga_2\,v^{1+(4/a)} \q\q\ {\rm and} \q\q 
8v_{ss}\=\ga_1\,v^{1+(8/a)}  \] 
admit  a common \so\ $v=s^{-a/4}$, i.e. 
\beq\lb{a/4} u(x,y)\=(x^2-y^2)^{-a/4} \eeq
with
\[\ga_1\=(a/2)(a+4) \q\q\ \ga_2\=-(a/4)(3a+4)\ .\]
The most interesting situation occurs when $a\not=\pm 4,\pm 8,\ldots$: in 
this case the \so\ (\ref{a/4}) exists only in  the region $|x|\ge |y|$. 

It can be noticed that \eq\ (\ref{pde2}) admits also one of the trivial \sys\ mentioned at the beginning of this paper (and discussed in \cite{ND}), precisely the scaling
\[ X'\=x\frac{\partial}{\partial x}+y\frac{\partial}{\partial y}- {a\ov 2}\,u\,
\frac{\partial}{\partial u}\]
and -- in addition -- that the above \so\ (\ref{a/4}) turns out to be invariant under this \sy\ (see also \cite{CPC}  for the reduction of Eq. (\ref{genpde}) obtained thanks to the  \sy\ $X'$). 
In other words, one can say that  \so\  (\ref{a/4}) exhibits the peculiar property of being
simultaneously  invariant under  the ``standard'' \sy\ $X'$ and the
weak conditional  one $Y$.

Taking now advantage from  the ``exceptional'' \sy\ $X$ [Eq.(\ref{exsy})] 
which is present in this case, we can construct from 
the above \so\ (\ref{a/4}) a continuous family of \so s to Eq. 
(\ref{pde2}) using  the rule 
(\ref{util}-5$'$): we obtain
\beq\lb{Dsh}u(x,y)\=\big[ x^2- \big(y+\la (x^2+y^2)\big)^2\big]^{-a/4}\eeq
which holds only in the interior of the two circles centered resp. in 
$x_0\!=\! 1/(2\la)$, $y_0\!=\!-1/(2\la)$ and $x_0\!=-1/(2\la)$, 
$y_0\!=\!-1/(2\la)$, both of radius $1/(\sqrt{2}\la)$, and excluding  
their intersection (it is not restrictive to assume $\la>0$). 

A \so\ of 
this form is relevant in the context of plasma physics because 
it describes a specially interesting ``double-D" shaped  toroidal plasma 
equilibrium which exhibits  a rather flat plasma profile with strong 
pressure  gradients  and current gradients at the edge (see \cite{CPC}
for some comment on this point). This is another aspect that enhances 
the  exceptional  character of the \sy\ $X$ considered in the present note
and the interest in it.

\sk\sk\noindent
{\bf Acknowledgments}. It is a pleasure to thank the referees for their valuable remarks, 
and in particular one of them for pointing out to me the two references [12-13].

\label{lastpage}


\begin{thebibliography}{99}
\small

\bibitem{ND}  Cicogna, G.: Symmetry classification of quasi-linear PDE's 
containing arbitrary functions, Nonlinear Dynamics  {\bf 51}, 309--316 
(2008) 

\bibitem{Tok}  Wesson, J.:$\!\,$  Tokamaks, The Oxford Engineering
Ser. 48, Oxford, Clarendon (1997)%, 2nd Edition  

\bibitem {FOG}  Frewer, M., Oberlack, M.,  Guenther, S.:  Symmetry
investigations on the incompressible stationary axisymmetric Euler \eq s
with swirl, Fluid Dyn. Research {\bf 39}, 647--664 (2007) 

\bibitem{CPC}   Cicogna, G.,  Pegoraro, F.,  Ceccherini, F.:  Symmetries, 
weak symmetries and related solutions of the Grad-Shafranov equation, 
Phys. of Plasmas {\bf  17},  102506-1/8 (2010)

\bibitem{Bl} Bluman, G.W. and Cole, J.D.: The general similarity \so\ of 
the heat \eq , {\it J. Math. Mech.} {\bf 18}, 1025--1042 (1969)

\bibitem{LW}   Levi, D. and  Winternitz, P.: Non-classical \sy\ 
reduction:  
example of the Boussinesq \eq , {\it J. Phys. A}  {\bf 22}, 
2915--2924  (1989) 

\bibitem{Fu}  Fushchych, W.I.  (Ed.), {\em Symmetry analysis of equations 
of
mathematical physics}, Inst. of Math. Acad. of Science of Ukraine, Kiev 
(1992), and  {\it Conditional symmetries of
the equations of mathematical physics}, in   Modern group analysis:
advanced analytical and computational methods in mathematical physics,
(Ibragimov N.H., Torrisi M. and Valenti A., Eds.), Kluwer, Dordrecht 
(1993)

\bibitem{Po} Kunzinger, M. and Popovych, R.O.: Is a nonclassical \sy\ a \sy
?, Proc. 4th Workshop ÒGroup Analysis of Differential Equations \& 
Integrable systemsÓ, pp. 107--120, Cyprus (2008); arXiv:0903.0821 (2009)

\bibitem{OR} Olver,  P.J. and Rosenau, Ph.:  The construction of special 
\so 
s to partial differential \eq s,  
 Phys. Lett. A   {\bf 114}, 107--112  (1986), and Group-invariant \so s of 
differential \eq s, 
 SIAM J. Appl. Math.  {\bf 47},  263--278 (1986)

\bibitem{CK}  Cicogna, G.: A discussion on the different notions of 
symmetry of differential equations, 
Proc. Inst. Math. N.A.S. Ukr.  {\bf 50}, 77--84, 2004

\bibitem{CL}  Cicogna, G. and Laino, M.: On the notion of conditional \sy\ 
of differential \eq s, Rev. Math. Phys.  {\bf  18},   1--18 (2006)

\bibitem{F1} Fushchych, W.I.: Conditional \sy\  of the \eq s of nonlinear mathematical physics, 
Ukrainian Math. J. {\bf 43}, No. 11, 1350--1364 (1991)

\bibitem{F2}  Fushchych, W.I. and Zhdanov, R.Z.: Anti-reduction of the nonlinear wave equation, 
Reports of N.A.S. of Ukraine  {\bf 11}, 37--41 (1993), translated in: Fushchych, W.I., 
Scientific Works  {\bf 5}, 116--119 (2003).


\end{thebibliography}
\end{document}